\documentclass{Jetpl}
\twocolumn

\lat

\title{Longitudinal and Transverse NMR in \\Superfluid $^3$He in Anisotropic Aerogel}

\rtitle{Longitudinal and Transverse NMR in Superfluid $^3$He in
Anisotropic Aerogel}

\sodtitle{Longitudinal and Transverse NMR in Superfluid $^3$He in
Anisotropic Aerogel}

\author{V.\,V. Dmitriev$^{+}$\thanks{e-mail: dmitriev@kapitza.ras.ru},
D.\,A. Krasnikhin$^+$, N. Mulders$^*$, V.\,V. Zavjalov$^+$, D.\,E.
Zmeev$^+$}

\rauthor{V.\,V. Dmitriev, D.\,A. Krasnikhin, N. Mulders, V.\,V.
Zavjalov, D.\,E. Zmeev}

\sodauthor{V.\,V. Dmitriev, D.\,A. Krasnikhin, N. Mulders, V.\,V.
Zavjalov, D.\,E. Zmeev}

\address{$^+$P.\,L. Kapitza Institute for Physical Problems RAS,
2 Kosygina str., 119334 Moscow, Russia\\~\\
$^*$Department of Physics and Astronomy, University of Delaware,
Newark, Delaware 19716, USA}

\dates{\today}{*}

\abstract{It was found that NMR properties of both superfluid
phases of $^3$He in anisotropic aerogel can be described in terms
of the bulk superfluid order parameters with the orbital order
parameter vector fixed by anisotropy of the aerogel sample. It was
also shown that by a proper squeezing it is possible to get the
aerogel sample with isotropic NMR properties.}


\begin{document}

\maketitle {\bf Introduction.} The system ``liquid $^3$He + high
porosity silica aerogel'' allows to investigate the influence of
disorder on \emph{p}-wave superfluidity. The disorder is
introduced by the aerogel strands. The diameter of the strands
($\sim$30\,\AA) is much less than the correlation length of the
bulk superfluid $^3$He and a characteristic distance between them
is large enough, so the superfluidity of $^3$He is not fully
suppressed \cite{Porto, Halperin}. In a weak magnetic field there
exist two superfluid phases of $^3$He in aerogel called A-like and
B-like \cite{Barker}. The A-like phase appears on cooling from the
normal phase at pressures above $\approx$\,20\,bar and exists in
rather large temperature range in a metastable (supercooled)
state. It is established \cite{Barker,Dmit1} that the B-like phase
is analogous to the B phase of ``usual'' bulk $^3$He, i.e. it is
described by the same Balian-Werthamer (BW) order parameter as the
bulk B phase. It is also known \cite{Barker} that the A-like phase
belongs to the family of Equal Spin Pairing (ESP) phases, but the
exact structure of its order parameter is still unclear. G.\,E.
Volovik suggested that the A-like phase in aerogel is described by
the Larkin-Imry-Ma (LIM) model with spatially random orientation
of Anderson-Brinkman-Morel (ABM) order parameter\cite{Volovik1,
Volovik2}. In the bulk A phase the order parameter is also
described by the ABM model, but it is spatially homogeneous.
I.\,A. Fomin has proposed the so called ``robust'' ESP phase
\t~the phase in which the orientation of the order parameter is
not influenced by the presence of aerogel \t~as a possible
candidate for the A-like phase \cite{Fomin1, Fomin2}. In previous
experiments in aerogel with porosity of about 98\% \cite{Barker,
Dmit2, Dmit3, Dmit4} (and with 97.5\% and 99.3\% porosity samples
\cite{Osheroff, Ishikawa1, Ishikawa}) it was found that the NMR
properties of the A-like phase are different from the properties
of the A phase of bulk $^3$He. In the same time the observed NMR
properties also do not correspond well to both LIM and ``robust''
phase models \cite{Dmit4}. Recent experiments with 98\% aerogel
\cite{Kunimatsu} have clarified the problem: it was found that in
squeezed by 1-2\,\% aerogel sample the A-like phase behaves as the
A phase of the bulk $^3$He with vector $\hat{\bf l}$ fixed along
the axis of deformation (i.e. along the axis of anisotropy). This
observation agrees with recent theoretical studies, where it was
shown that even for small anisotropy ($\sim$1\,\%) spatially
homogeneous A phase order parameter is more favorable than LIM or
``robust'' state \cite{Volovik3, Fomin3}. Consequently, if the
sample is inside a glass tube as probably was in \cite{Barker,
Ishikawa} (or there is no large enough gap between the sample and
epoxy walls of the cell or spacers fixing the sample as it was in
\cite{Dmit2, Dmit3, Dmit4}) then a difference in thermal
contraction coefficients of aerogel and the walls could result in
uncontrolled deformation and complicate interpretation of the
results.

Here we present and compare the results of our recent NMR studies
of the A-like and the B-like phases in 3 aerogel samples. As it is
shown below two of them were anisotropic, while the third one had
isotropic NMR properties.

{\bf Experimental details.} Experiments were done at pressures of
26.0\,bar and 28.6\,bar in the magnetic fields range of 40-528\,Oe
(corresponding to NMR frequencies from 132 to 1714\,kHz). We used
98.2\,\% porosity aerogel in which silica strands occupy only
about 1.8\,\% of the whole volume. Three experimental cells
(similar to that described in \cite{Dmit1, Dmit3}) with three
different aerogel samples were used. The samples had a cylindrical
form (sample 1: diameter=4\,mm, height=3.5\,mm; samples 2 and 3:
diameter=5\,mm, height=1.5\,mm) with the axis oriented along
\textbf{z}. Samples 1 and 2 were laying freely inside the epoxy
cells, so that there were large enough gaps ($\approx$\,0.15\,mm)
between the sample and the side and top walls. Correspondingly we
believe that no additional deformation could appear during
cooldown from room temperature due to thermal shrinkage of the
cell, which is expected to be about 1\,\%. Sample 3 was fixed in
the cell by 4 paper spacers (0.15\,mm thick, width=0.5\,mm and the
length along \textbf{z}-axis is 1.5\,mm). The spacers were glued
to the side walls and the aerogel sample was presumably squeezed
by them in the \textbf{x}-\textbf{y} plane after cooldown from
room temperature.

The cells were surrounded by transverse NMR coils with their axes
oriented along the \textbf{x} direction. Standard NMR setup was
used, i.e. radiofrequency (RF) excitation was applied to the NMR
circuit; the voltage across the coil was amplified by a
preamplifier and then detected by lock-in amplifier (in case of
continuous wave, CW, NMR) or by digital oscilloscope (in case of
pulsed NMR). Cell with sample 1 also had a superconducting
longitudinal NMR coil for the longitudinal resonance experiments.
The corresponding NMR circuit was cold and had the fixed frequency
(9095\,Hz) with the quality factor of 1860. External steady
magnetic field \textbf{H} could be rotated in the
\textbf{z}-\textbf{y} plane: most of the experiments were done for
\textbf{H}$\parallel$\textbf{z} (longitudinal field) and for
\textbf{H}$\perp$\textbf{z} (transverse field).

The temperature was obtained by copper nuclear demagnetization
refrigerator and was measured with a vibrating wire viscometer and
a quartz tuning fork situated in a large volume connected to the
experimental cell by a short ($\approx$\,5\,mm) and narrow
(diameter of 1\,mm) channel. In order to avoid signal from
paramagnetic solid $^3$He on the surface of aerogel strands, all
our aerogel samples were preplated with $\sim$2.5 atomic layers of
$^4$He. Consequently no Curie-Weiss behavior of spin
susceptibility was observed in our experiments.

\begin{figure}
\begin{centering}
\includegraphics[width=1.0\linewidth]{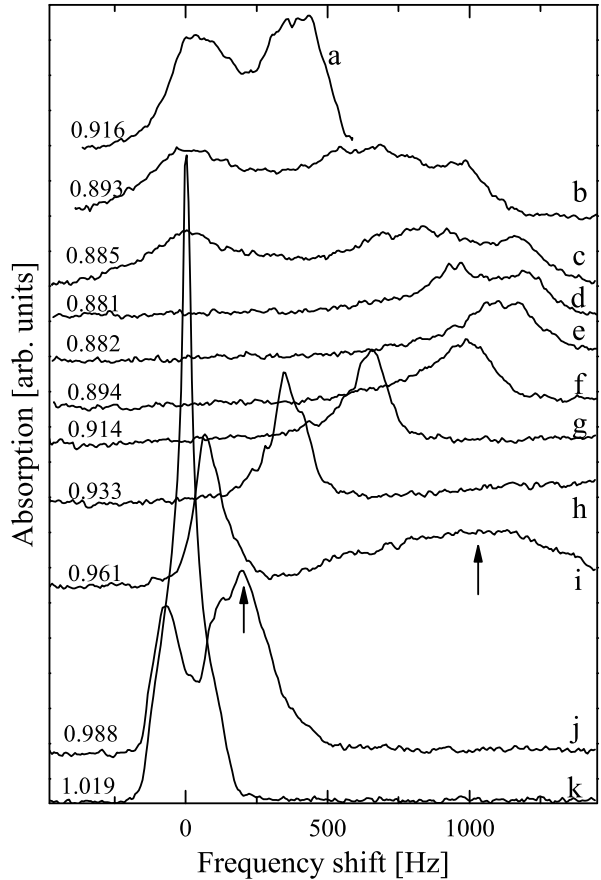}
\caption  {Fig.1. CW NMR lines in sample~1 in transverse field
(\textbf{H}$\perp$\textbf{z}) obtained by cooling from the normal
phase down to onset of the A-like$\rightarrow$B-like transition
and subsequent warming (for clarity the lines are shifted in
y-direction). Temperatures in units T/T$_{ca}$ are shown near each
NMR line, where T$_{ca}$ is the superfluid transition temperature
of $^3$He in aerogel. P=26.0\,bar, H=58.3\,Oe,
T$_{ca}$=0.80\,T$_c$, where T$_c$ is the superfluid transition
temperature of bulk $^3$He}
\end{centering}
\end{figure}

{\bf CW NMR experiments in samples 1 and 2.} For
\textbf{H}$\parallel$\textbf{z} it was found that in samples 1 and
2 CW NMR line in the A-like phase had large negative frequency
shift from the Larmor value. The value of the shift was of the
same order as in \cite{Kunimatsu}. We also observed that the
negative shift converts to positive as the direction of the
external magnetic field is changed to \textbf{H}$\perp$\textbf{z}.
It is known that in the ABM phase the frequency shift from the
Larmor value ($\Delta\omega$) depends on the angle $\xi$ between
\textbf{H} and the orbital vector $\hat{\bf l}$:
\begin{equation}
\Delta\omega=-\frac{\Omega_A^2}{2\omega}\cos(2\xi), \label{shift}
\end{equation}
where $\omega$ is the NMR frequency and $\Omega_A$ is the Leggett
frequency. Accordingly our observations can be explained if we
suggest that the A-like phase in samples 1 and 2 corresponds to
the ABM model and both these samples are intrinsically anisotropic
with the main axis of anisotropy directed along \textbf{z}. It was
also found that the anisotropy of sample 1 was not homogeneous:
for \textbf{H}$\perp$\textbf{z} at low enough temperature CW NMR
line was rather broad and had 3 distinct maxima (line ``c'' in
Fig.1). The observed maxima can be attributed to 3 parts of the
sample where $\xi$ is approximately equal to 90$^\circ$ (i.e.
$\hat{\bf l}\parallel$\textbf{z}), 70$^\circ$ and 45$^\circ$
correspondingly ($\Delta\omega$ is maximal at $\xi$=90$^\circ$ and
equals zero at $\xi$=45$^\circ$). The temperature width of the
transition from the A-like to B-like phase in sample 1 was rather
wide ($\sim$0.02\,T$_{ca}$) and it was found that at first the
transition occurred for the less shifted part of the A-like phase
NMR line (see lines c,d,e in Fig.1). This allowed us to cool the
sample down to the A-like$\rightarrow$B-like transition region and
then warm up so that the A-like phase survived only in part of the
sample (lines e-j in Fig.1) and the other part was in the B-phase
(the B-like phase signal has much larger frequency shift and in
Fig.1 it can be seen only near T$_{ca}$ as shown by arrows near
lines i and j). The obtained in such a way NMR lines in the A-like
phase were rather narrow and we used them for the quantitative
measurements assuming that they correspond to the part of sample 1
where $\hat{\bf l}\parallel$\textbf{z} and $\xi$=90$^\circ$ (or
$\xi$=0$^\circ$ in the case of longitudinal orientation of
\textbf{H}). In particular, when we rotated \textbf{H} at a fixed
temperature from transverse to longitudinal orientation
(\textbf{H}$\parallel$\textbf{z}) the shift of the A-like phase
line changed the sign, but the absolute value of the shift
remained the same as it is expected from (\ref{shift}). The
obtained dependencies of the frequency shift in the A-like phase
on temperature may be recalculated to $\Omega_A^2$ (open symbols
in Fig.2).

In sample 1 we also carried out longitudinal NMR experiments. In
these experiments we were sweeping the temperature while recording
the signal from the longitudinal NMR coil. To simplify the
interpretation we describe below the results where the A-like
phase was left only in the region of the sample where $\hat{\bf
l}\parallel$\textbf{z}. For the ABM order parameter the frequency
of longitudinal NMR $\Omega_{\parallel}$ should depend on $\xi$:
\begin{equation}
\Omega_{\parallel}=\Omega_{A}\sin(\xi). \label{long}
\end{equation}
The axis of our longitudinal coil was oriented along \textbf{z}
(or along $\hat{\bf l}$) so we were not able to see longitudinal
NMR signal for \textbf{H}$\perp$\textbf{z}. Therefore we used
angles $\xi$=0$^\circ$ and 60$^\circ$ between \textbf{H} and
$\hat{\bf l}$ (or \textbf{z}). In accordance with (\ref{long}) no
longitudinal NMR signal in the A-like phase was found for
$\xi$=0$^\circ$ while for $\xi$=60$^\circ$ the longitudinal NMR
signal was clearly seen. The obtained relationship between the
transverse resonant frequency and the longitudinal resonant
frequency recalculated to the case \textbf{H}$\perp\hat{\bf l}$
was found to well correspond to the ABM phase:
\begin{equation}
\Delta\omega=(0.52\pm0.04)\frac{\Omega_{\parallel}^2}{\omega}
\end{equation}

The vector $\hat{\bf l}$ in the B-like phase is also oriented by
the anisotropy \cite{Kunimatsu,Elbs}. For
\textbf{H}$\parallel$\textbf{z} the B-like phase line had only
positive frequency shift with a sharp peak near the Larmor value
(which corresponds to $\hat{\bf l}\parallel$\textbf{H} for the BW
order parameter) and there was no problem to distinguish it from
the A-like phase signal. For \textbf{H}$\perp$\textbf{z} the
signal from the B-like phase had large positive frequency shift
and interfered with the A-like phase signal only close to the
superfluid transition temperature in aerogel $T_{ca}$ (see lines i
and j in Fig.1). Sometimes after the A-like$\rightarrow$B-like
transition the high frequency end of the B-like phase NMR line had
a narrow sharp peak. Following \cite{Hakonen, Dmit5} we attributed
this peak to the textural defect in which vector \textbf{n} of the
BW order parameter is perpendicular to \textbf{H}. The
corresponding frequency shift can be recalculated to the Leggett
frequency of the B-like phase (black circles in Fig.2). Note that
the frequency shifts in the A-like and B-like phases depend on
temperature in a different way: on warming the frequency shift in
the A-like phase drops to zero at the temperature about
0.02\,T$_{ca}$ lower than in the B-like phase. This result shows
that at least near T$_{ca}$ the A-like phase does not exactly
correspond to the spatially homogeneous ABM phase.

\begin{figure}
\begin{centering}
\includegraphics[width=1.0\linewidth]{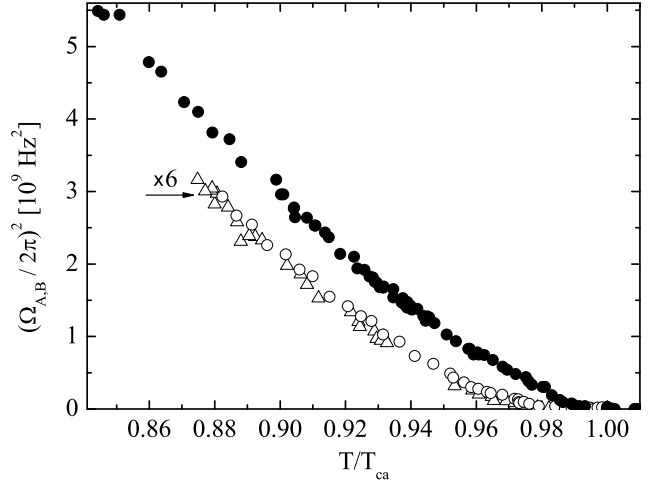}
\caption{Fig.2. Leggett frequencies $\Omega_B$ and $\Omega_A$in
sample 1 calculated from transverse NMR frequency shift assuming
$\hat{\bf l}\parallel$\textbf{z}. {\Large $\bullet$} - the B-like
phase; $\triangle$,{\Large $\circ$} - the A-like phase, calculated
from positive and negative frequency shift data correspondingly.
Note that $\Omega_A^2$ is multiplied by a factor of 6.
T$_{ca}$=0.80\,T$_c$, P=26.0\,bar}
\end{centering}
\end{figure}

{\bf Pulsed NMR experiments in the A-like phase.} The results of
pulsed NMR experiments is another argument that allows us to
conclude that the A-like phase in anisotropic aerogel has the same
order parameter as in the bulk A phase. The pulsed NMR experiments
were performed in $^3$He in sample 2. CW NMR properties in this
sample were the same as in sample 1 the only difference being that
the NMR line had only one broad peak, which implies that
anisotropy in sample 2 was more homogeneous. Unfortunately we were
not able to perform pulsed NMR in longitudinal field due to the
bulk B phase in the gaps between the aerogel sample and the cell
walls. This signal was close to the Larmor value and in the
observed total free induction decay signal (FIDS) interfered with
the signal from the A-like phase. In transverse field the signal
from the B phase was shifted far enough from the A-like phase
signal, so that we were able to apply long enough ($\sim$0.2\,ms)
and intense RF pulses to tip magnetization only in the A-like
phase. Note that in this orientation of \textbf{H} the situation
is similar to the case of bulk $^3$He-A, where $\hat{\bf l}$ is
fixed perpendicular to \textbf{H} by spin-orbital interaction. The
results of pulsed NMR experiments in the A-like phase in
transverse field are shown in Fig.3. The obtained dependence of
the FIDS frequency shift $\Delta\omega$ on the magnetization
tipping angle~$\beta$ can be described as

\begin{equation}\label{pulsed}
\Delta\omega=A\left(\frac{1+3\cos\beta}{4}\right)=\frac{\Omega_A^2}{8\omega}\left(1+3\cos\beta\right).
\end{equation}
This dependence is characteristic of the ABM order parameter
\cite{BS} and was observed earlier in the bulk A phase \cite{CO}.

\begin{figure}
\begin{centering}
\includegraphics[width=1.0\linewidth]{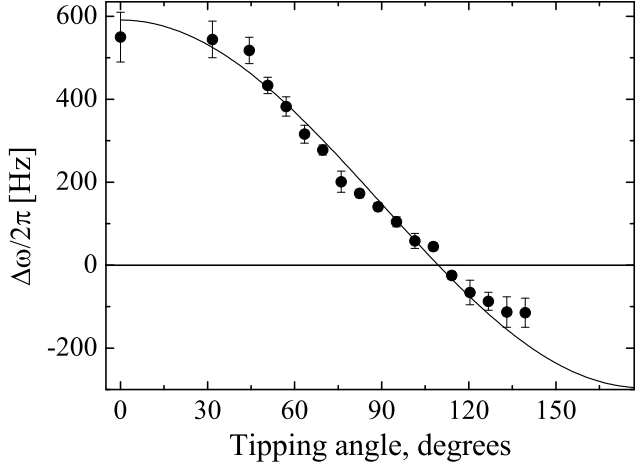}
\caption{Fig.3. Initial frequency shift of FIDS versus tipping
angle $\beta$ in the A-like phase in sample 2. The point at
$\beta=0^\circ$ is from CW NMR measurements. Solid line is the
best fit of expression~(\ref{pulsed}) to the data with
$A/2\pi$=590\,Hz. H=97.3\,Oe, P=28.6 bar, T=0.933\,T$_{ca}$,
T$_{ca}$=0.82\,T$_c$}
\end{centering}
\end{figure}

\begin{figure}
\begin{centering}
\includegraphics[width=1.0\linewidth]{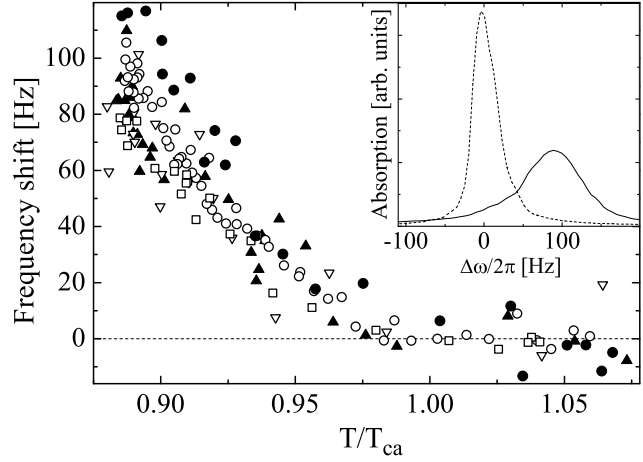}
\caption{Fig.4. The frequency shift (the first moment of the CW
NMR absorption line) in the A-like phase versus temperature in
sample 3 for different angles $\psi$ between \textbf{H} and
\textbf{z}. {\Large $\circ$} - $\psi$=0$^\circ$, $\square$ -
$\psi$=15$^\circ$, {\Large $\bullet$} - $\psi$=31$^\circ$,
$\bigtriangledown$ - $\psi$=51$^\circ$, $\blacktriangle$ -
$\psi$=60$^\circ$. H=142\,Oe, P=26.0\,bar.  Insert: CW NMR
absorption lines for $\psi$=0$^\circ$. Dashed line -
T=1.09\,T$_{ca}$, solid line - T=0.89\,T$_{ca}$ }
\end{centering}
\end{figure}

\begin{figure}
\begin{centering}
\includegraphics[width=1.0\linewidth]{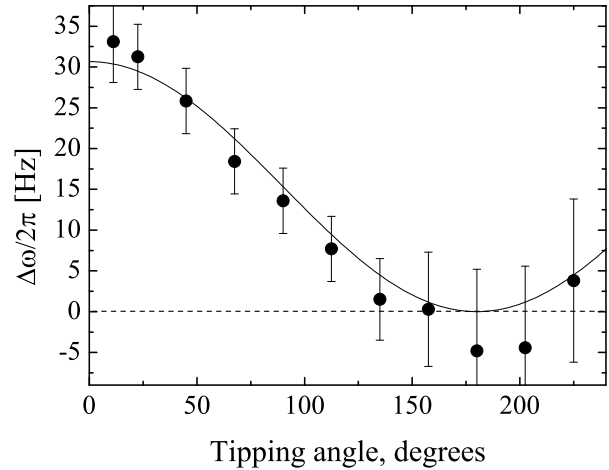}
\caption{Fig.5. Initial frequency shift of FIDS versus tipping
angle $\beta$ in the A-like phase in sample 3. Solid line is the
best fit of the data by  $A(1+\cos\beta)$ with $A$=15.4\,Hz.
H=528\,Oe, P=26.0 bar, T=0.89\,T$_{ca}$}
\end{centering}
\end{figure}

{\bf NMR experiments in sample 3.} Samples 2 and 3 were cut from
the same aerogel rod with a high-speed diamond wheel saw. However
NMR properties of sample 3 in the A-like phase were quite
different from the properties of sample 2. We found that in sample
3 mean frequency shift of the NMR absorption line (the first
moment of the line) was positive and did not depend on the angle
$\psi$ between \textbf{z} and \textbf{H} (see Fig.4). The value of
the shift (recalculated to the same conditions) was 5-6 times
smaller than the absolute value of the shift in the A-like phase
in sample 2 for transverse and longitudinal orientations. It is
possible that after cooling from room temperature sample 3 was
compressed in the \textbf{x}-\textbf{y} plane by the side wall
spacers. Computer simulations of 1\,\% squeezing by our spacers
show that in $\sim$60\,\% of the sample the increase of the
density in transverse plane should be in the range of 0.4 - 1\,\%.
This compression may compensate possible intrinsic longitudinal
anisotropy and two situations may realize. The first: the
anisotropy axis is deflected from \textbf{z} by some angle which
is different in different parts of the sample due to the
inhomogeneity of compression. Correspondingly $\xi$ and
$\Delta\omega$ should essentially vary over the sample. The
second: in the main part of the sample the compression decreases
the intrinsic longitudinal anisotropy below some critical value,
so that some state with isotropic NMR properties (e.g. LIM or
``robust'') becomes more favorable. The first situation probably
was realized in \cite{Osheroff} where the observed NMR line in the
A-like phase was very broad and where the mean frequency shift
depended on orientation of \textbf{H}. In our case the second
situation is more probable because CW NMR line in the A-like phase
had positive frequency shift for any $\psi$ and was rather narrow
(see the insert in Fig.4).

We also have done pulsed NMR experiments in sample 3 for
$\psi$=0$^\circ$. It was found that the frequency shift of the
FIDS from the Larmor value is approximately proportional to
$(1+\cos\beta)$ (Fig.5). We should note that the same dependence
of the FIDS frequency on the tipping angle as in Fig.5 was
observed in \cite{Ishikawa} in 97.5\% aerogel where in CW NMR
experiments the frequency shift in the A-like phase was also
positive and small \cite{Ishikawa1, Ishikawa} and of the same
order as in sample 3. Small and positive frequency shift in the
A-like phase in CW NMR was observed also in another aerogel sample
\cite{Dmit4}. The problem of identification of the order parameter
of the A-like phase in these cases remains unsolved.

{\bf Conclusions.} We have confirmed that the behavior of the
A-like and B-like states of superfluid $^3$He in anisotropic
aerogel can be described in terms of the corresponding bulk order
parameters (ABM and BW) assuming that the orbital order parameter
vector $\hat{\bf l}$ in both phases is fixed along the anisotropy
axis. The same statement was reported in \cite{Kunimatsu}. Here is
the summary of our observations indicating such a behavior.
\begin{enumerate}
\item The dependence of the FIDS on the magnetization tipping
angle in the A-like phase is the same as in the A phase of bulk
superfluid. \item The relationship between the frequencies of
longitudinal and transverse resonances in the A-like phase is the
same as in the bulk A phase. At low enough temperatures the ratio
of the Leggett frequencies in the A-like and B-like phases is
close to that in bulk superfluid.  \item The frequency shift in
the A-like phase changes from negative to positive as the magnetic
field \textbf{H} is rotated from longitudinal to transverse
orientation with respect to the anisotropy axis. The absolute
value of the shift is the same in these two orientations. No
longitudinal resonance is observed in the longitudinal
orientation. \item The CW NMR signal in the B-like phase has a
large peak near the Larmor frequency in longitudinal orientation
of \textbf{H} and is significantly shifted in the transverse
orientation of \textbf{H}.
\end{enumerate}

Nevertheless the observed behavior of the A-like phase in
anisotropic aerogel at temperatures close to T$_{ca}$ is not
completely described by the model of spatially homogeneous ABM
order parameter: it remains unclear why on warming the frequency
shift in the A-like phase drops to zero at essentially lower
temperature than in the B-like phase. We also have observed that
the A-like$\rightarrow$B-like transition temperature depends on
the NMR frequency shift. The same observation was reported in
\cite{Kunimatsu}, but in our experiments the transition at first
occurred in parts of the sample where the absolute values of the
frequency shift are smaller, that is opposite to the results of
\cite{Kunimatsu}.

We should note that in the described experiments the A-like phase
was obtained by cooling from the normal phase without any
additional external perturbations. However it was found that if
during cooldown through T$_{ca}$ we apply a set of RF tipping
pulses (even as small as 12$^\circ$) then NMR properties of the
obtained A-like phase change: the shift of the CW NMR line
essentially decreases and the frequency shift of FIDS in pulsed
NMR becomes proportional to $\cos\beta$ \cite{we}.

The results obtained in sample 3 show that the anisotropy may be
compensated by squeezing the aerogel in proper direction. The NMR
properties of the obtained state can not be described by the model
of homogeneous ABM phase or by smooth variation of orientation of
$\hat{\bf l}$ of the ABM phase over macroscopic distance. At the
same time our results do not exclude the states with isotropic NMR
properties: the Larkin-Imry-Ma state of the ABM phase, in which
the orientational long-range order is destroyed at distances
smaller than the dipole length; and the ``robust'' state. At the
moment we cannot distinguish between the two states.

{\bf Acknowledgements.} We thank I.\,A. Fomin and G.\,E. Volovik
for useful discussions and comments. The research was supported by
the Russian Foundation for Basic Research (06-02-17185), the
Ministry of Education and Science of Russia (NSh-9725.2006.2) and
CRDF (RUP1-MO-04-2632).


\begin{thebibliography}{99}


\bibitem{Porto}
J.\,V. Porto and J.\,M. Parpia, Phys. Rev. Lett. {\bf 74}, 4667
(1995).

\bibitem{Halperin} D.\,T. Sprague, T.\,M. Haard, J.\,B. Kycia, et al.,
Phys. Rev. Lett. {\bf 75}, 661 (1995).

\bibitem{Barker}
B.\,I. Barker, Y. Lee, L. Polukhina, et al., Phys.~Rev.~Lett. {\bf
85}, 2148 (2000).

\bibitem{Dmit1}
V.\,V. Dmitriev, V.\,V. Zavjalov, I.\,V. Kosarev, et al., Pis'ma v
ZhETF {\bf 76}, 371 (2002) [JETP~Lett. {\bf 76}, 321 (2002)].

\bibitem{Volovik1}
G.\,E. Volovik, Pis'ma v ZhETF {\bf 63}, 281 (1996) [JETP~Lett.
{\bf 63}, 301 (1996)].

\bibitem{Volovik2}
G.\,E. Volovik, Pis'ma v ZhETF {\bf 84}, 533 (2006) [JETP~Lett.
{\bf 84}, 455 (2006)].

\bibitem{Fomin1}
I.\,A. Fomin, Pis'ma v ZhETF {\bf 77}, 285 (2003) [JETP~Lett. {\bf
77}, 240 (2003)].

\bibitem{Fomin2}
I.\,A. Fomin, J. of Low Temp. Phys. {\bf 134}, 769 (2004).

\bibitem{Dmit2}
V.\,V. Dmitriev, I.\,V. Kosarev, N. Mulders, et al., Physica B
{\bf 329-333}, 296 (2003).

\bibitem{Dmit3}
V.\,V. Dmitriev, I.\,V. Kosarev, N. Mulders, et al., Physica B
{\bf 329-333}, 320 (2003).

\bibitem{Dmit4}
V.\,V. Dmitriev, L.\,V. Levitin, N. Mulders, D.\,E. Zmeev, Pis'ma
v ZhETF {\bf 84}, 539 (2006) [JETP~Lett. {\bf 84}, 461 (2006)];
arXiv:cond-mat/0607789.

\bibitem{Osheroff}
J.\,E. Baumgardner, D.\,D. Osheroff, Phys. Rev. Lett. {\bf 93},
155301 (2004).

\bibitem{Ishikawa1}
H. Nakagawa, K. Obara, H. Yano et al., J. of Low Temp. Phys. {\bf
138}, 159 (2005).

\bibitem{Ishikawa}
O. Ishikawa, R. Kado, H. Nakagawa et al., AIP~Conf.~Proc. {\bf
850}, 235 (2006).

\bibitem{Kunimatsu}
T. Kunimatsu, T. Sato, K. Izumina et al., Pis'ma v ZhETF {\bf 86},
244 (2007) [JETP~Lett. {\bf 86} (2007)]; arXiv:cond-mat/0612007

\bibitem{Volovik3}
G.\,E. Volovik, arXiv:0704.2484; Submitted to J. of Low Temp.
Phys. (Proceedings of the Symposium on Quantum Fluids and Solids,
QFS2007, Kazan, 1-6 August, 2007).

\bibitem{Fomin3}
I.\,A. Fomin, arXiv:0707.4222; Submitted to J. of Low Temp. Phys.
(Proceedings of the Symposium on Quantum Fluids and Solids,
QFS2007, Kazan, 1-6 August, 2007).

\bibitem{Elbs}
J. Elbs, E. Collin, Yu.\,M. Bunkov et al., arXiv:0707.3544.

\bibitem{Hakonen}
P.\,J. Hakonen, M. Krusius, M.\,M. Salomaa et al., J. of Low Temp.
Phys. {\bf 76}, 225 (1989).

\bibitem{Dmit5}
V.\,V. Dmitriev, N. Mulders, V.\,V. Zavjalov, D.\,E. Zmeev, AIP
Conf. Proc. {\bf 850}, 225 (2006).

\bibitem{BS}
W.\,F. Brinkman, H. Smith, Phys. Lett. {\bf 51A}, 449 (1975).

\bibitem{CO}
L.\,R. Corruccini, D.\,D. Osheroff, Phys. Lett. {\bf 51A}, 447
(1975); Phys. Rev. B {\bf 17}, 126 (1978).

\bibitem{we}
V.\,V. Dmitriev, D.\,A. Krasnikhin, N. Mulders, D.\,E. Zmeev,
arXiv:0709.3006; Submitted to J. of Low Temp. Phys. (Proceedings
of the Symposium on Quantum Fluids and Solids, QFS2007, Kazan, 1-6
August, 2007).
\end{thebibliography}
\end{document}